\documentclass[fleqn,10pt]{arXiv}
\pdfoutput=1

\usepackage[english]{babel}
\usepackage{natbib}
\usepackage{float}

\definecolor{color1}{RGB}{120,0,0}
\definecolor{color2}{RGB}{0,20,20}
\definecolor{color3}{RGB}{0,0,0}

\usepackage{hyperref}
\hypersetup{hidelinks,colorlinks,breaklinks=true,urlcolor=color2,citecolor=color1,
            linkcolor=color1,bookmarksopen=false,pdftitle={Title},pdfauthor={Author}}

\JournalInfo{arXiv preprint}
\Archive{}

\PaperTitle{No evidence of fish biodiversity effects on coral reef ecosystem functioning across scales \\ \vspace{0.25cm} \normalsize{Comment on ``Tropical fish diversity enhances coral reef functioning across multiple scales'' by Lefcheck et al.}}
\PaperSubtitle{No evidence of fish biodiversity effects on coral reef ecosystem functioning}

\Authors{Tarik C. Gouhier\textsuperscript{1,*} and Pradeep Pillai\textsuperscript{1}}
\affiliation{\textsuperscript{1}\textit{Marine Science Center, Northeastern University, 430 Nahant Road, Nahant, MA 01908}}
\affiliation{*\textbf{Corresponding author}: tarik.gouhier@gmail.com}

\newcommand{\keywordname}{Keywords}

\Abstract{We demonstrate that the conclusions drawn by \citet{lefcheck2019} regarding the
positive effects of fish diversity on coral reef ecosystem functioning across scales
are flawed because of a series of conceptual and statistical issues that include
spurious correlations, the conflation of population size and species diversity effects,
and a failure to recognize that observing a biodiversity effect at multiple sites is
not equivalent to observing it at multiple scales.}

\begin{document}
\flushbottom
\maketitle
\thispagestyle{firstpage}


\section*{Introduction}

\citet{lefcheck2019} sought to show the beneficial effects of tropical fish biodiversity
on coral reef ecosystem functioning at multiple scales. To do so, they collected data from
video and transect surveys at ten sites to determine whether $\alpha$ and $\beta$ species diversity of
fish led to an increase in ecosystem functioning in the form of higher grazing rates at multiple
scales and whether grazing rates enhanced ecosystem structure by reducing turf abundance and
promoting coral abundance. Below, we describe a number of major conceptual and statistical flaws
in their study that undermine their results and conclusions.

\begin{figure*}[!htb]\centering
\includegraphics[width=\linewidth]{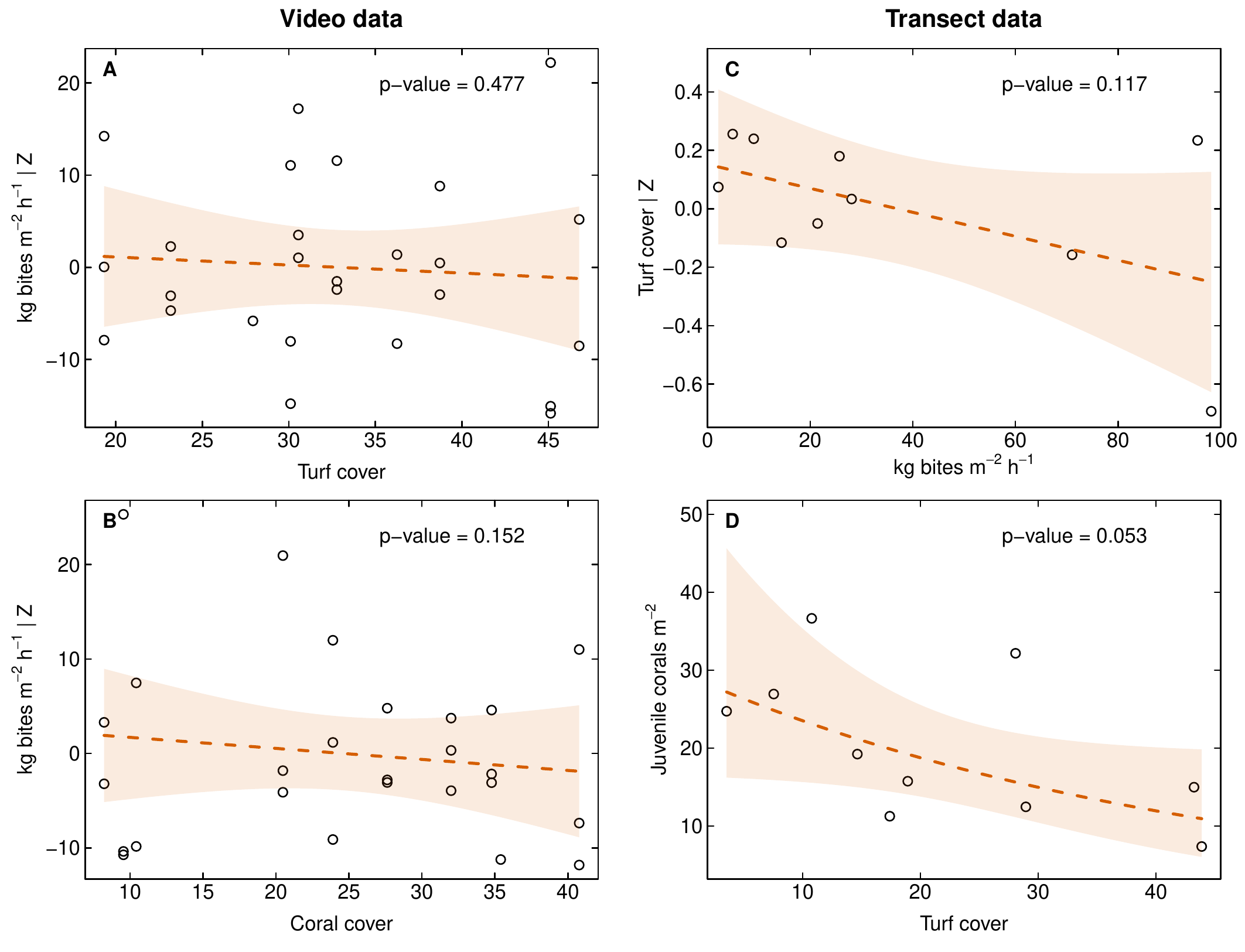}
\caption{\textbf{No relationship between mass-standardized bite rate and ecosystem functioning}. No relationship between mass-standardized bite rate
and turf cover (A) or coral cover (B) in the video dataset. No relationship between mass-standardized bite rate and turf cover (C) or between turf
cover and juvenile coral density (D) in the transect dataset.}
\label{fig1}
\end{figure*}

\section*{Conceptual and statistical flaws}

\subsection*{Grazing rate is an inappropriate measure of ecosystem functioning}
The first conceptual issue lies in the use of the grazing rate (bite rate) as a measure of ecosystem functioning. Increased grazing of turf can
promote ecosystem functioning if it reduces turf abundance and thus leads to an increase in reef-building corals via competitive release. Ecosystem
functioning should thus be measured in terms of the impact of grazing on turf or coral rather than its rate. Grazing rate, on its own, is merely one
of an infinite number of community or system-level properties that could also be arbitrarily designated as an ``ecosystem function''. Its relevance
arises only due to its potential impact on an ecosystem property of interest. Showing a significant positive relationship between various diversity
metrics and grazing rate is thus necessary but not sufficient to demonstrate greater ecosystem functioning.

However, the mixed-effects model in Lefcheck et al. showed no relationship between mass-standardized bite rate and either turf cover or coral cover
in the video dataset (Fig. \ref{fig1}A,B). The lack of a relationship suggests that changes in the mass-standardized bite rate do not
translate into changes in turf or coral cover. Additionally, no relationship emerges when the mass-standardized bite rate is regressed against
turf cover rather than turf height in the transect dataset (Fig. \ref{fig1}C). Similarly, no relationship exists when regressing juvenile coral
recruitment against turf cover (Fig. \ref{fig1}D). This means that the entire case for a meaningful effect of bite rate on ecosystem functioning
emerges only when turf height (not cover) is used in one of the two datasets. The results that link mass-standardized bite rate
and ecosystem functioning are thus not robust.

\begin{figure*}[!htb]\centering
\includegraphics[width=\linewidth]{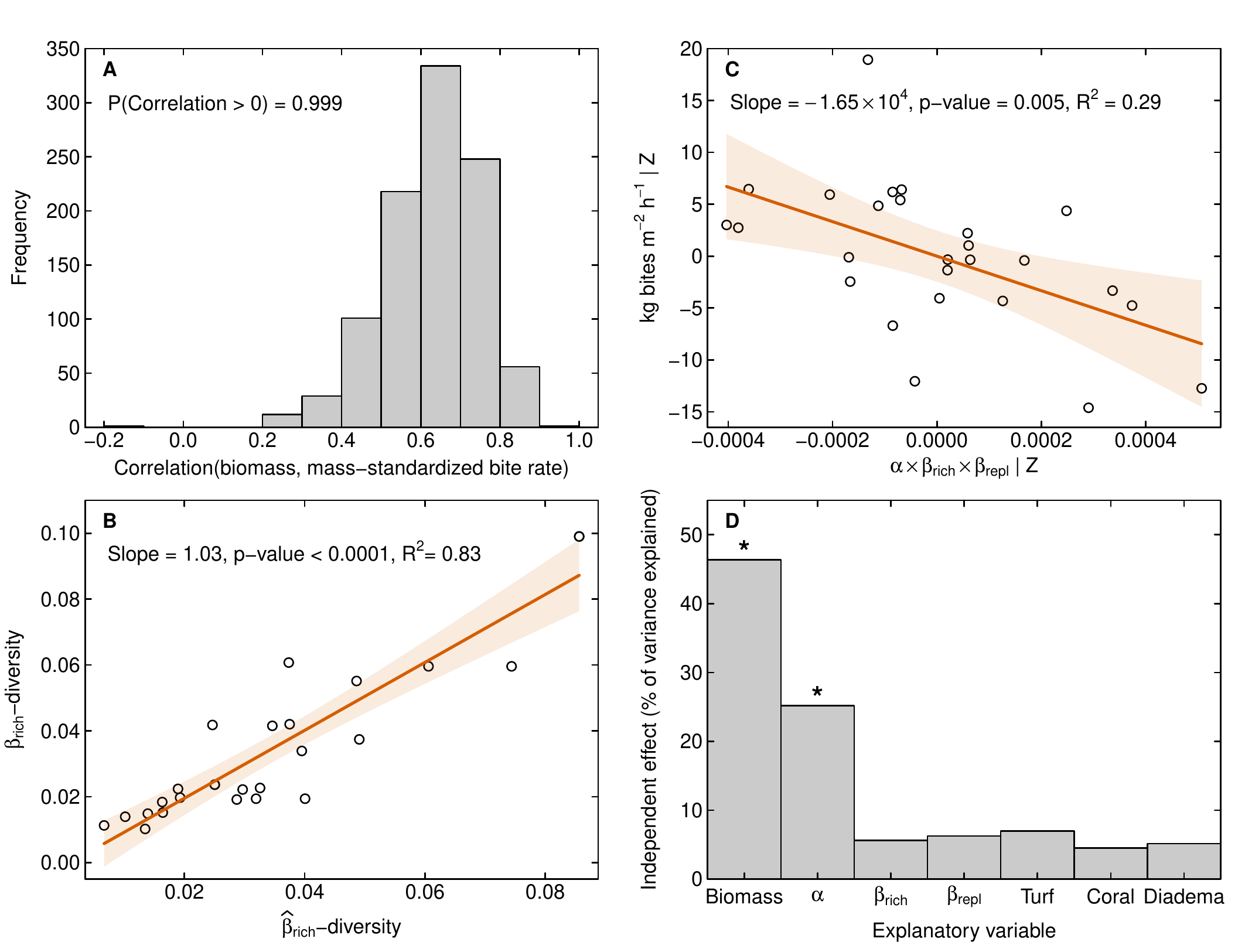}
\caption{\textbf{Statistical issues with attribution of mass-standardized bite rate}. (A) Distribution of spurious correlations induced between
biomass and mass-standardized bite rate when biomass and bite rate are independent random variables drawn from a uniform distribution across 1000
Monte Carlo simulations. (B) Collinearity between diversity metrics shown via a significant positive relationship between $\beta_\text{rich}$
observed and $\hat{\beta}_\text{rich}$ predicted from a multiple regression of $\beta_\text{rich}$  against $\alpha$ and $\beta_\text{rich}$
explaining 83\% of the variance. (C) Significant negative partial effect of the three-way interaction between $\alpha$, $\beta_\text{rich}$  and
$\beta_\text{repl}$ diversity on the mass-standardized bite rate. (D) The independent effect of each explanatory variable as a percentage of the
variance explained in the mass-standardized bite rate. The model explains 89\% of the total variance. Asterisks indicate statistically significant
variables ($\text{p-value} < 0.05$).}
\label{fig2}
\end{figure*}

\subsection*{No evidence that diversity promotes ecosystem functioning across scales}

The second conceptual issue stems from the claim that diversity promotes ecosystem functioning across scales and the suggestion that the results
presented in Lefcheck et al. are consistent with the spatial insurance hypothesis. Spatial insurance effects occur when ecosystem functioning is
enhanced and more stable at the regional scale as a result of local sites undergoing favorable conditions rescuing those undergoing unfavorable ones
\citep{loreau2003}. Because spatial insurance effects emerge at the regional scale, they cannot be detected by regressing local ecosystem functioning
against local factors such as diversity and biomass at all sites. Although the results presented in Lefcheck et al. show that localized measures of
diversity promote local ecosystem functioning, this local-scale effect of diversity was misinterpreted as evidence of a multi-scale effect because it
arose at multiple sites. However, observing a biodiversity effect at multiple sites is not the same as observing it at multiple scales. Here, the
suggestion that diversity enhances ecosystem functioning across multiple scales is simply not supported by the data. Additionally, the conceptual
link drawn to the spatial insurance hypothesis does not make sense given that the analyses were all performed at the local scale.

\subsection*{Inevitable relationships between biomass, diversity and bite rate}
The mixed-effects model presented in Lefcheck et al. showed a significant positive relationship between mass-standardized bite rate and biomass.
However, this relationship is at least partially attributable to a spurious correlation because the response variable is the bite rate scaled by the
explanatory variable (biomass). Monte Carlo simulations show that this leads to spurious correlations between mass-standardized bite rate and biomass
when biomass and bite rate are independent random variables drawn from a uniform distribution (Fig. \ref{fig2}A). When the spurious correlation issue
is fixed by using the non mass-standardized bite rate as a response variable, the positive effect of biomass remains significant but that is because
biomass is acting as a surrogate for the number of fish observed at each site (correlation = 0.97, $\text{p-value} < 0.0001$). Hence, a significant
positive relationship between biomass and bite rate was to be expected since increasing the number of fish leads to both greater total biomass and a
larger number of total bites.

A similar issue arises with $\alpha$ diversity (local species richness), which was also positively associated with bite rate. This was interpreted as
a local diversity effect, with more species yielding a higher total bite rate, perhaps because of complementarity in resource use between fish
species. However, the relationship between total bite rate and $\alpha$ diversity was bound to be positive since increasing $\alpha$ diversity is
largely tantamount to increasing the total number of fish as long as the community is not saturated (i.e., no zero-sum game whereby the addition of
an individual from one species leads to the loss of an individual from another species). Because increasing species richness leads to an increase in
the number of fish (correlation = 0.54, $\text{p-value} = 0.007$), and adding individual fish will increase the total number of bites, the
relationship between total bite rate and $\alpha$ diversity essentially has to be positive. Hence, since $\alpha$ diversity is at least partially
acting as a surrogate for the total number of fish, it is not surprising to see a positive relationship emerge between total bite rate and $\alpha$
diversity. However, this is likely due to a population size effect rather than a true species diversity effect. Indeed, when the effect of population
size is controlled for by first regressing the mass-standardized bite rate against the number of fish, the residuals of the mass-standardized bite
rate are unrelated to either biomass $(\text{p-value} = 0.49)$ or $\alpha$ diversity $(\text{p-value} = 0.2)$.

\subsection*{Spurious relationship between $\beta_\text{rich}$ and bite rate}
Lefcheck et al. found a positive relationship between local bite rate and $\beta_\text{rich}$ diversity---a measure of the uniqueness of the community
at a given site---and claimed that it represented evidence of a spatial insurance effect. As mentioned above, this local relationship between
$\beta_\text{rich}$  diversity and bite rate cannot represent evidence of a regional spatial insurance effect. Additionally, there is no clear
mechanism by which higher $\beta_\text{rich}$ diversity can lead to a higher bite rate at the local scale, as the positive effects of
$\beta_\text{rich}$ diversity on bite rate can only emerge when sites are aggregated at larger spatial scales. Any positive effect of a site’s
compositional uniqueness expressed via $\beta_\text{rich}$  diversity on local bite rate would be captured by local factors such as $\alpha$
diversity and biomass. It is more likely that the positive effect of $\beta_\text{rich}$ diversity on local bite rate reported in Lefcheck et al. is
due to multicollinearity between the explanatory variables $\alpha$, $\beta_\text{repl}$  and $\beta_\text{rich}$  diversity included in their mixed-
effects model (Fig. \ref{fig2}B).

Furthermore, perhaps under the mistaken impression that the additive partitions of total $\beta$ diversity---namely $\beta_\text{repl}$ and
$\beta_\text{rich}$---had to be orthogonal, the authors verified that all possible two-way interactions between $\alpha$ diversity and the components
of $\beta$ diversity were not significant but failed to test and include the significant three-way interaction between $\alpha$,
$\beta_\text{repl}$  and $\beta_\text{rich}$  diversity in their model ($\text{p-value} = 0.03$). The coefficient associated with this significant
three-way interaction is negative, so an increase in any of the three diversity metrics will lead to a reduction in the mass-standardized bite rate
(Fig. \ref{fig2}C). Standard statistical practice dictates that in the presence of such a significant negative three-way interaction, the positive
main effects of $\alpha$ and $\beta_\text{rich}$ diversity should not be interpreted because their independent effects on mass-standardized bite rate
are not consistent \citep{whitlock2008,quinn2002,sokal2011,zar1999}. Hence, the positive main effects of $\alpha$ and $\beta_\text{rich}$ diversity
that constitute the backbone of Lefcheck et al.’s conclusions are suspect at best. To verify this claim, we used hierarchical partitioning
\citep{chevan1991,macnally2000} to determine the independent effect of each explanatory variable on mass-standardized bite rate and found that only
biomass and $\alpha$ diversity were significant and collectively represented 72\% of the variance explained, whereas $\beta_\text{rich}$  and
$\beta_\text{repl}$ diversity were not significant and collectively represented only 12\% of the variance explained (Fig. \ref{fig2}D). This is not
surprising since the positive effects of $\beta$ diversity cannot emerge at the local scale.

\subsection*{Misinterpreted evidence for complementary}
Multiple regression was used to relate mass-standardized bite rate at each site to the proportional biomass of each species in order to determine
whether there was ‘complementarity’ between species in terms of their contributions to bite rate. However, there seems to be some confusion about how
to interpret these results. Significant positive relationships between proportional abundance and mass-standardized bite rates across sites cannot be
interpreted as evidence of ‘complementarity’ without ensuring that the bite rates observed in multi-species communities at the very least exceed
those expected based on the bite rates observed in their constituent single-species populations. Otherwise, significant relationships between
proportional biomass and bite rates could just as likely arise because of redundancy between species that equally contribute to the bite rate at all
sites.

If anything, these significant relationships provide potential evidence for a lack of ‘complementarity’ at the regional scale. Indeed, a significant
relationship between bite rate and a focal species' proportional biomass means that the focal species contributes significantly to the bite rate
across all sites. Hence, fewer significant relationships indicates fewer species ‘dominating’ or contributing consistently to the local bite rate
across all sites and suggests greater spatial ‘complementarity’, with some species contributing more to the local bite rate at a subset of sites. In
this case, the proportional biomass of four of the nine species was significantly related to bite rate across all sites, which suggests that about
56\% (5/9) of species are spatially ‘dominant’ and about 44\% (4/9) of species either contribute differentially (‘complementarily’) or not at all to
the local bite rate across sites.

\subsection*{Conclusion}

Overall, we believe that the conceptual and statistical issues outlined above demonstrate that there is no evidence that fish diversity promotes
ecosystem functioning across scales. Establishing this important result would require linking $\alpha$ and $\beta$ diversity to greater ecosystem
functioning in the form of higher coral cover or lower turf cover beyond the local scale by aggregating the data across sites as other researchers
have done in terrestrial systems \citep{winfree2018}.

\phantomsection
\section*{Acknowledgments}

We acknowledge support from the National Science Foundation (OCE-1458150, CCF‐1442728).

\phantomsection
\bibliographystyle{ecology}
\bibliography{preprints}

\newcommand{\noopsort}[1]{}
\begin{thebibliography}{9}
\expandafter\ifx\csname natexlab\endcsname\relax\def\natexlab#1{#1}\fi
\expandafter\ifx\csname url\endcsname\relax
  \def\url#1{{\tt #1}}\fi
\expandafter\ifx\csname urlprefix\endcsname\relax\def\urlprefix{URL }\fi

\bibitem[{Chevan and Sutherland(1991)}]{chevan1991}
Chevan, A. and M.~Sutherland. 1991.
\newblock Hierarchical partitioning.
\newblock American Statistician, {\bf 45}:90--96.

\bibitem[{Lefcheck et~al.(2019)Lefcheck, {Innes-Gold}, Brandl, Steneck, Torres,
  and Rasher}]{lefcheck2019}
Lefcheck, J.~S., A.~A. {Innes-Gold}, S.~J. Brandl, R.~S. Steneck, R.~E. Torres,
  and D.~B. Rasher. 2019.
\newblock Tropical fish diversity enhances coral reef functioning across
  multiple scales.
\newblock Science Advances, {\bf 5}:eaav6420.

\bibitem[{Loreau et~al.(2003)Loreau, Mouquet, and Gonzalez}]{loreau2003}
Loreau, M., N.~Mouquet, and A.~Gonzalez. 2003.
\newblock Biodiversity as spatial insurance in heterogeneous landscapes.
\newblock PNAS, {\bf 100}:12765--12770.

\bibitem[{Mac~Nally(2000)}]{macnally2000}
Mac~Nally, R. 2000.
\newblock Regression and model-building in conservation biology, biogeography
  and ecology: {{The}} distinction between and reconciliation of 'predictive'
  and 'explanatory' models.
\newblock Biodiversity and Conservation, {\bf 9}:655--671.

\bibitem[{Quinn and Keough(2002)}]{quinn2002}
Quinn, G.~P. and M.~J. Keough. 2002.
\newblock Experimental {{Design}} and {{Data Analysis}} for {{Biologists}}.
\newblock {Cambridge University Press}, first edition.

\bibitem[{Sokal and Rohlf(2011)}]{sokal2011}
Sokal, R.~R. and F.~J. Rohlf. 2011.
\newblock Biometry.
\newblock {W. H. Freeman}, New York, fourth edition.

\bibitem[{Whitlock and Schluter(2008)}]{whitlock2008}
Whitlock, M.~C. and D.~Schluter. 2008.
\newblock The {{Analysis}} of {{Biological Data}}.
\newblock {Roberts and Company Publishers}, first edition.

\bibitem[{Winfree et~al.(2018)Winfree, Reilly, Bartomeus, Cariveau, Williams,
  and Gibbs}]{winfree2018}
Winfree, R., J.~R. Reilly, I.~Bartomeus, D.~P. Cariveau, N.~M. Williams, and
  J.~Gibbs. 2018.
\newblock Species turnover promotes the importance of bee diversity for crop
  pollination at regional scales.
\newblock Science, {\bf 359}:791--793.

\bibitem[{Zar(1999)}]{zar1999}
Zar, J.~H. 1999.
\newblock Biostatistical {{Analysis}}.
\newblock {Prentice-Hall, Inc.}, Upper Saddle River, NJ., fourth edition.

\end{thebibliography}


\end{document}